\documentclass[prd, aps, showpacs]{revtex4}
\usepackage[latin1]{inputenc} 

\usepackage[spanish,activeacute]{babel}

\def\be{\begin{equation}}
\def\te{\end{equation}}
\def\ee{\end{equation}}
\def\ba{\begin{eqnarray}}
\def\bea{\begin{eqnarray}}
\def\nn{\nonumber\\}
\def\tea{\end{eqnarray}}
\def\ea{\end{eqnarray}}
\def\eea{\end{eqnarray}}

\begin{document}

\title{Non abelian hydrodynamics and heavy ion collisions}

\author{E. Calzetta}
\email{calzetta@df.uba.ar}
\affiliation{Departamento de F\'\i sica, Facultad de Ciencias Exactas y Naturales, Universidad de Buenos Aires and IFIBA, CONICET, Ciudad Universitaria, Buenos Aires 1428, Argentina}

\begin{abstract}
The goal of the relativistic heavy ion collisions (RHIC) program is to create a state of matter where color degrees of freedom are deconfined. The dynamics of matter in this state, in spite of the complexities of quantum chromodynamics, is largely determined by the conservation laws of energy momentum and color currents. Therefore it is possible to describe its main features in hydrodynamic terms, the very short color neutralization time notwithstanding. In this lecture we shall give a simple derivation of the hydrodynamics of a color charged fluid, by generalizing the usual derivation of hydrodynamics from kinetic theory to the non abelian case.
\end{abstract}
\maketitle

\section{Introduction} 
Non abelian gauge theories enjoy the property of asymptotic freedom whereby coupling strengths get weaker at higher energy and density \cite{nobel}. It therefore seems possible, by increasing the energy and density of a matter droplet, to go beyond confinement to a new state of matter where color degrees of freedom may reveal themselves. To create this state of matter is the goal of the relativistic heavy ion collisions (RHIC) program  \cite{BRAHMS05,PHOBOS05,STAR05,PHENIX05}. By colliding two heavy ions against each other at relativistic speeds, a central region is created where conditions approach those at the Big Bang (at least, more closely than ever before on Earth) \cite{bang}. The matter in this region expands, cools and eventually breaks up into hadrons \cite{Ruus86}. The number and spatial distributions of the created particles carry relics of their hot and dense past. 

The theoretical description of a heavy ion collision is usually split into three eras, an early one of formation, a second of expansion and cooling, and finally the third of hadron formation, (almost) free flight and detection \cite{BelLan56,BJOR83}. The transitions from first to second and from second to third are open problems on their own. There is general agreement that the second stage may be described by hydrodynamics \cite{Risc98,Hir08,Hei10,CalHu08}. The smoking gun is the so-called elliptic flow, which shows that pressure anisotropy in the early stage is being converted into particle yield anisotropy in the late state \cite{Roma09}. This kind of transmutation is the hallmark of hydrodynamic models, and it is hard to obtain from alternative approaches. The fluid undergoing hydrodynamic evolution is the so-called quark gluon plasma (QGP).

In this lecture we shall argue that at least the latest times of the first stage, and some phenomena pertaining to the second, such as energy deposition from a fast colored particle crossing the QGP \cite{jets}, may also be described within a hydrodynamic framework, namely one where color currents appear explicitly \cite{Kaj80,manuelrev,ManMor06,BBK11,color}. It must be noted that if hydrodynamics is regarded as the theory of slow degrees of freedom, on time scales larger than the relaxation towards local thermal equilibrium, then it becomes an issue whether a hydrodynamics of colored particles has any range of applicability: color will be neutralized on very short time scales \cite{whitening}. However, we do not regard hydrodynamics in this way, but rather as the simplest effective theory consistent with known conservation laws.

This perspective shift however poses its own problems, because usually hydrodynamics is derived from the more fundamental kinetic theory description under the assumption that the relaxation time is indeed the shortest time scale available. To dispose of that assumption we must build a new framework for the derivation of hydrodynamics.

In this Lecture we shall first go through the derivation of hydrodynamics from kinetic theory for a fluid with no conserved charges, and then use that discussion as a blueprint for the formulation of non abelian hydrodynamics.

Our starting point is relativistic kinetic theory as formulated by Israel and others \cite{Isr72,Isr88}. The problem is how to reduce it to the hydrodynamic level. In the non relativistic case, the method of choice is the so-called Chapman - Enskog expansion, which leads to the Navier - Stokes equations \cite{CHACO39}. In the relativistic case, it works likewise. However, in this case the former success becomes troublesome, because the relativistic Navier-Stokes equation is known to have deep causality and stability problems \cite{HisLin83,Ols90,OlsHis90}, to be discussed below.

Fortunately there is a second standard approach to hydrodynamics, namely the so-called Grad expansion, which disposes of the causality and stability issue. The Grad approach is based on the principle that a successful description of the evolution of a non ideal fluid must incorporate degrees of freedom other than temperature, fugacity and fluid velocity. The extra degrees of freedom vanish identically for an ideal fluid, but follow dynamical equations of their own in the non ideal case.

The Grad approach links the kinetic and hydrodynamic levels by solving the closure problem (namely, the problem of identifying the one particle distribution function at the kinetic level which relates to a given fluid configuration at the hydrodynamic level) through a quadratic ansatz for the one particle distribution function. This leads to a one particle distribution function which is negative in some regions in momentum space. This unphysical property underlies the fact that the Grad expansion is only asymptotic, and holds in the limit when the relaxation time is much shorter than any other time scale in the problem. It may be observed that this is more than just a formal issue, since this drawback contaminates the predictions of the theory regarding observables which are sensitive to high momenta.

The enlarging of the degrees of freedom is a common feature in several formulations of real relativistic hydrodynamics. In many approaches, such as the so-called  Israel - Stewart theory \cite{is1,is2} or Extended Thermodynamics \cite{extended} the extra variables are the components of the viscous energy-momentum tensor (VEMT) $\Pi_{\mu\nu}$ itself. The dynamical equations for $\Pi_{\mu\nu}$ have been derived in a number of ways, such as carefully taking moments of the kinetic equation \cite{DMNR12}, a systematic gradient expansion of the kinetic theory \cite{BRW06}, from AdS-CFT correspondence \cite{cft} or simply writing down all terms consistent with the symmetries of the theory up to a certain order \cite{BRSSS}. We shall call these theories ``second order fluid dynamics'' (SOFD) for short; the presentation in \cite{EXG10} is a suitable representative.

In the Geroch - Lindblom Divergence type theories \cite{dtt}, on the other hand, the new variables do not have a direct physical interpretation; $\Pi_{\mu\nu}$ may depend nonlinearly on them.  These theories can be rigorously proven to be free of causality and stability problems, but their  physical foundations remain elusive \cite{dtt2}. Extended thermodynamics, on the other hand, is presented as an expansion in powers of different relaxation times \cite{Diego}, the large order behavior of this expansion being unknown.

We shall therefore adopt a different point of view in the formulation of a generalized Grad hydrodynamics \cite{CalPR10,CalPR12,Cal13}. The basic insight is that any realistic kinetic theory displays a whole hierarchy of relaxation times, from the relaxation times characteristic of large scale inhomogeneities and anisotropies to the much shorter relaxation times of hard modes. Fluctuations in the hydrodynamic modes are the so-called soft modes. They relax on time scales which may be long with respect to other relevant parameters. The slowly varying soft modes  are perceived by the harder modes as externally imposed thermodynamic forces which prevent their relaxation to true equilibrium. In other words, the hard modes relax not to equilibrium but to a nonequilibrium steady state constrained by the instantaneous configuration of the soft modes. It has been known from long ago that such steady states are the solutions to variational problems \cite{Ono61,Jay80}. Prigogine among others has proposed that they are the extrema of the entropy \emph{production}, as opposed to the extrema of the entropy itself, which are the true equilibria \cite{Pri55}. Known proofs of the so-called ``Prigogine theorem'' are restricted to linear irreversible thermodynamics \cite{Lan75,Bru06}; we shall appeal to it on a heuristic rather than formal basis. In short, the idea is that most hard modes are in the linear regime most of the time anyway, so a theory which is good in the linear regime is good enough to compute global observables such as stress tensor components, but we shall  not attempt to formulate this insight in any rigorous way \cite{Kli91,Kli95}.

This lecture is organized as follows. Next Section presents the usual derivations of hydrodynamics from kinetic theory for neutral conformal particles. The presentation is an abridged version of our recent review \cite{Cal13}. In the following Section we apply this methodology to the case of a color charged fluid interacting with classical non abelian fields. We shall not be concerned with the derivation of the kinetic theory for this second case, it may be obtained either from the so-called Wong equations for point color charges \cite{Kaj80,manuelrev} or else from quantum field theory \cite{CalHu08}. The discussion in this third Section follows \cite{color}. We conclude with some very brief final remarks.

\section{Hydrodynamics from kinetic theory}
In the kinetic theory description the transport equation reads 
\be
p^{\mu}\partial_{\mu}f=\frac{-1}{\tau}\mathrm{sign}\left(p^0\right)I_{col}
\label{kinetic}
\te
$f=f\left(x,p \right)$ is the one particle distribution function depending on event $x$ and momentum $p$, $\tau$ is the so-called relaxation time and $I_{col}$ is the collision integral to be discussed below. The  currents are the particle (charge) current
\be
J^{\mu}=e\int\;Dp\;p^{\mu}f
\te
and the energy-momentum tensor

\be
T^{\mu\nu}=\int\;Dp\;p^{\mu}p^{\nu}f
\te

where 
\be
Dp =\frac{2d^4p\delta\left(p^2\right)}{\left(2\pi\right)^3}=\frac{d^4p}{\left(2\pi\right)^3 p}\left(\delta\left(p^0-p \right) + \delta\left(p^0+p \right)\right) 
\te
For simplicity we assume massless particles.

In this section we do not assume particle number conservation. To enforce energy-momentum conservation  we require
\be
\int\;Dp\;p^{\mu}\mathrm{sign}\left(p^0\right)I_{col}=0
\te
The equilibria are the thermal distributions
\be
f_0=\exp\left\{-\left|\beta_{\mu}p^{\mu}\right|\right\}
\te
where $\beta_{\mu}=u_{\mu}/T$ is the inverse temperature four vector. The velocity $u^{\mu}$ is a normalized ($u^ 2=-1$, we use  $-+++$ signature for the Minkowsky metric) time like four vector, while $T$ is the temperature as measured in the local rest frame. We define the energy density $\rho$ and the velocity from the Landau prescription: $u^{\mu}$ is the only time like eigenvector of the energy momentum tensor, and $-\rho$ is the corresponding eigenvalue: $T^{\mu}_{\nu}u^{\nu}=-\rho u^{\mu}$ \cite{LL6}. For a given $T^{\mu\nu}$ we can always find a  local equilibrium distribution  $f_0$ with the same energy density and four velocity as from the Landau-Lifshitz prescription. We then define a temperature from $\rho=\sigma T^4$. The energy momentum tensor built from $f_0$ takes the form 

\be 
T_0^{\mu\nu}=\rho u^{\mu}u^ {\nu}+p\Delta^{\mu\nu}
\te 
where $p=\rho/3$ is the pressure and
$\Delta^{\mu\nu}=\eta^{\mu\nu}+u^{\mu}u^ {\nu}$ is the projector orthogonal to the velocity; $\eta^{\mu\nu}$ is the Minkowsky metric.
It follows that the viscous energy momentum tensor (VEMT)

\be 
\Pi^{\mu\nu}=T^{\mu\nu}-T_0^{\mu\nu}
\label{VEMT}
\te 
is traceless and transverse
\be 
\Pi^{\mu\nu}u_{\nu}=0
\label{TTC}
\te 
We parametrize

\be 
f=f_0\left[1+Z\right]
\label{param}
\te 
with $Z=0$ at equilibrium. The transversality condition becomes

\be
\int\;Dp\;p^{\mu}\left(-u_{\nu}p^{\nu}\right)f_0Z=0
\label{trans}
\te
We assume a simple Boltzmann type entropy flux 

\be
S^{\mu}=-\int\;Dp\;\left(\mathrm{sign}\left(p^0 \right)\right)p^{\mu}f\left[\ln \frac{f}{f_0}-1\right]
\te
This is really the relative entropy with respect to $f_0$ \cite{Sag13}; when $f_0$ is thermal it is rather the flux of the Massieu function $-F/T$, where $F$ is the free energy; however we shall call it entropy for short. This understood, we get the entropy production

\be
S^{\mu}_{,\mu}=\frac{1}{\tau}\int\;Dp\;I_{col}\ln\left[1+Z\right] 
\te
The goal of hydrodynamics is to replace the function $f$ by a finite number of fields (including at least temperature and velocity) in terms of which we could express the particle number, energy - momentum and entropy currents, and to derive the equations of motion for these fields. The equations of motion must be covariant and causal, and thermal states are expected to be stable. Given a suitable set of hydrodynamic fields, the idea is to solve the closure problem, that is, to express $f$ (or $Z$, cfr. eq. (\ref{param})) as a functional of the hydrodynamic fields, whereby the currents may be computed by performing the relevant integrals, and the equations of motion are derived from the kinetic equation eq. (\ref{kinetic}).

\subsection{The Chapman-Enskog approach}
In the Chapman-Enskog approach the hydrodynamic fields are the same as in a thermal state, namely just $\beta_{\mu}$ for neutral conformal particles. 
The Chapman-Enskog procedure is to solve the closure problem by obtaining a formal expansion for $Z$ (eq. (\ref{param})) in powers of $\tau$. To this end, one parametrizes

\be
Z=\tau Z_1+\tau^2Z_2+\ldots
\te
The ``spatial'' derivatives $\Delta^{\mu\nu}T_{,\nu}$ and $\Delta^{\mu\nu}u^{\lambda}_{,\nu}$ are regarded as zeroth order quantities, while the ``time'' derivatives $\dot{T}=u^{\mu}T_{,\mu}$ and
$\dot{u}^{\lambda}=u^{\mu}u^{\lambda}_{,\mu}$ are derived from energy-momentum conservation, which is also a necessary consistency condition. 

To find $Z_1$ we only need the linearized collision integral. This must be a symmetric operator, it must obey the energy momentum conservation constraint, must lead to non negative entropy production and must admit thermal distributions as the only homogeneous solutions.

We adopt the Anderson - Witting prescription \cite{AndWit74,TakInu10}
\bea 
&&I_{col}\left( p\right) =\left| -u_{\mu}p^{\mu}\right|   f_0\left( p\right) \left[Z\left(p \right) \right.\nn
&-&\left.K_{\rho\sigma}p^{\rho}\mathrm{sign}\left(p^0\right)\int\;Dp'\;p'^{\sigma}\left(-u_{\mu}p'^{\mu}\right)  f_0\left( p'\right) Z\left( p'\right) \right] 
\tea

\be
K_{\rho\sigma}=\frac1A_3\left[-u_{\rho}u_{\sigma}+3\Delta_{\rho\sigma}\right]
\te

\be
A_k=\int\;Dp'\;\left| -u_{\mu}p^{\mu}\right|^k    f_0\left( p'\right)
\te

If $Z$ satisfies the constraint eq. (\ref{trans}), then the second term is zero and the entropy production is positive, provided $Z\ge -1$.
To first order
\be
Z_1=\frac{-1}{2T\left|p^{\alpha}u_{\alpha}\right|}\sigma_{\mu\nu}{p^{\mu}p^{\nu}}
\label{trans2}
\te
where $\sigma_{\mu\nu}$ is the shear tensor

\be 
\sigma_{\mu\nu}=\Delta^{\rho}_{\mu}\Delta^{\sigma}_{\nu}\left\lbrace u_{\rho;\sigma}+u_{\sigma;\rho}-\frac 23\Delta_{\rho\sigma}u^{\lambda}_{;\lambda}\right\rbrace 
\te 
The viscous energy momentum tensor is
\be 
\Pi_1^{\mu\nu} =-\eta\sigma^{\mu\nu}
\label{VEMTCE1}
\te 
where
\be
\eta=\frac{8}{5\pi^2}\tau T^4
\te
is the shear viscosity. Since we are assuming a conformal theory there is no bulk viscosity. The conservation equations 
\be
T^{\mu\nu}_{0,\nu}+\Pi^{\mu\nu}_{,\nu}=0
\te
split into an equation for the energy density

\be
\dot{\rho}+\left(\rho+p\right)u^{\nu}_{,\nu}+\frac12\Pi^{\mu\nu}\sigma_{\mu\nu}=0
\label{cons}
\te
where $\dot{\rho}=u^{\mu}\rho_{;\mu}$ and another one for the velocity
\be
\dot{u}^{\mu}+\Delta^{\mu\nu}\frac{p_{,\nu}+\Pi^{\lambda}_{\nu,\lambda}}{\left(\rho+p\right)}=0
\label{NSE}
\te
which under eq. (\ref{VEMTCE1}) becomes a covariant Navier-Stokes equation.

\subsubsection{The stability problem}
Eqs. (\ref{cons}) and (\ref{NSE}) fail the test of stability of equilibrium solutions. To show this, let us consider the linear perturbations around a solution with $u^{\mu}=\left(u^0,u\right)=$ constant. Let us  take $u$ in the $z$ direction. We shall consider plane waves propagating in the $z$ direction with frequency $\omega$ and wave number $K$. Since the theory is covariant, we may always perform a Lorentz transformation to the rest frame. Let $\omega'$ and $K'$ be the rest frame frequency and wave number. Write
\bea 
\delta' T'=T\delta' e^{-i\omega' t'+ik'_jx'^j} \nn
\delta' u'^i=\delta'^ i e^{-i\omega' t'+ik'_jx'^j}
\tea
The equations of motion become ($\gamma =\eta/\rho +p$)
\bea
-i\omega'\delta'+ic^2K'\delta'^{3}&=&0\nn
-i\omega'\delta'^{3}+iK'\delta'+\frac 43{\gamma}K'^2\delta'^{3} &=&0\nn
-i\omega'\delta'^{1,2}+{\gamma}K'^2\delta'^{1,2} &=&0
\tea

We observe that because  the imaginary parts of $\omega$ and $\omega'$ have the same sign we can check stability by checking the imaginary part of $\omega'$.	For a transverse perturbation we must have $\omega'=-i\gamma K'^2$. 
Under a Galilean transformation $K'=K$ and $\omega'=\omega-uK$, so the theory is stable in all frames. However, under a relativistic transformation $K'=\left(  K-u\omega'\right) /u^0$. If $u\not= 0$ we get a quadratic equation for $\omega'$
\be 
\omega'^2-2\left(\frac Ku+\frac{iu^{02}}{2\gamma u^2} \right)\omega' +\frac{K^2}{u^2}=0
\te 
When $u\to 0$, one solution is the expected one $\omega'=-i\gamma K^ 2$ but the second solution becomes $\omega'=iu^{02}/\gamma u^2$ and is unstable. For a longitudinal perturbation 
a similar analysis  leads to a cubic equation. When $u\to 0$ two roots correspond to damped sound waves; the third root $\omega'\approx 3i/4\gamma u^2$ is unstable.

\subsection{The Grad approach}
The Grad procedure takes the different strategy of keeping the form eq. (\ref{trans2}), but replacing $\sigma_{\mu\nu}$ by a new tensor $C_{\mu\nu}$ regarded as a new independent variable

\be
Z_G=\frac{-1}{2T\left|-p^{\alpha}u_{\alpha}\right|}C_{\mu\nu}{p^{\mu}p^{\nu}}
\te
$C_{\mu\nu}$ is defined up to  a multiple of $\eta_{\mu\nu}$, so we may assume it is traceless, and because of the constraint eq. (\ref{trans}) it must be transverse. 
The viscous energy momentum tensor becomes
\be 
\Pi_G^{\mu\nu}=-\frac{\eta}{\tau}C^{\mu\nu}
\label{VEMTG}
\te
The next step is to substitute this into the Boltzmann equation. The nonlocal term vanishes and we get 
\be
\frac{1}{\tau}\left(p^{\nu}u_{\nu}\right)f_0Z=\frac{\partial}{\partial x^{\mu}}\left[p^{\mu}f_0\left(1+Z \right) \right]
\label{BE2}
\te
To extract from there an equation for $C^{\mu\nu}$ we consider the moments of this equation. The zeroth moment vanishes and the first moment gives back energy - momentum conservation.
We get an equation for $C^{\mu\nu}$ from the second moments
\bea 
\frac {-1}{T^5}\frac{\partial}{\partial x^{\mu}}{T^{5}}u^{\mu}C_{\tau\sigma}&-&\Delta_{\sigma\lambda}C^{\rho\lambda\mu}\frac{\partial u_{\tau}u_{\rho}}{\partial x^{\mu}}
-\Delta_{\tau\rho}C^{\rho\lambda\mu}\frac{\partial u_{\sigma}u_{\lambda}}{\partial x^{\mu}}\nn
&+&\frac13\Delta_{\tau\sigma}C^{\rho\lambda\mu}\frac{\partial u_{\rho}u_{\lambda}}{\partial x^{\mu}}+\frac13\Delta_{\rho\lambda}C^{\rho\lambda\mu}\frac{\partial u_{\sigma}u_{\tau}}{\partial x^{\mu}}\nn
&=&\frac{1}{\tau}C_{\tau\sigma}
\label{GE}
\tea 
\bea 
C^{\rho\lambda\mu}&=& \left\{3u^{\rho}u^{\lambda}u^{\mu}
+ \Delta^{\rho\lambda}u^{\mu}+\Delta^{\mu\lambda}u^{\rho}+\Delta^{\mu\rho}u^{\lambda}\right.\nn
&-&\left.u^{\lambda}C^{\rho\mu}- u^{\rho}C^{\mu\lambda}-u^{\mu}C^{\rho\lambda} \right\}
\tea
For linear deviations from equilibrium
\be
\dot{C}_{\tau\sigma}+\frac{1}{\tau}C_{\tau\sigma}-\sigma_{\tau\sigma}+\mathrm{ho}=0
\te 
in the rest frame. This is an equation of the so-called Maxwell-Cattaneo type \cite{Max67,JosPre89}.

\subsubsection{Causality}
We wish to check that the Grad approach is consistent with causality and stability. As we did for the Chapman - Enskog approach, we consider a perturbation of an equilibrium state with non vanishing velocity. Introducing a new perturbation  for $C^{ij}$ we get the rest frame equations

\bea
-i\omega'\delta'+ic^2k'_j\delta'^j&=&0\nn
-i\omega'\delta'^i+ik'^i\delta' -i\frac{\gamma}{\tau}k'_jC'^{ij}&=&0\nn
-i\omega'C'^{ij}+\frac{1}{\tau}C'^{ij}-i\left( k'^i\delta'^j+k'^j\delta'^i-\frac 23\delta^{ij}k'_k\delta'^k\right) &=&0
\tea
We revert to the Chapman - Enskog equations with the replacement $\gamma\to\gamma /1-i\omega'\tau$. 

For transverse perturbations we now get a quadratic equation

\be 
\left(1-i\omega'\tau\right)\omega'+i\gamma K'^2=0
\te 
It is easy to see that both roots are stable when $u=0$. For general $u$ stability obtains if $\tau\ge\gamma$, while for longitudinal waves stability requires $\tau\ge 2\gamma$. With the expressions above $\tau\approx 5\gamma$.

\subsection{Entropy production variational method (EPVM)}
Although the Grad approach solves the stability issue, it has the lingering drawback that the suggested approximation for $f$ is not nonnegative definite. This leads to unphysical predictions for observables sensitive to large momenta.

To obtain a better closure we observe that on time scales short with respect to $\tau$ the fluid relaxes to a steady nonequilibrium state characterized by a non vanishing viscous energy momentum tensor; the relaxation to true equilibrium is a much slower process. We shall investigate the structure of the transient nonequilibrium state by applying the so-called Prigogine theorem, namely, nonequilibrium steady states are extrema of the entropy \emph{production}, constrained by the actual value of the VEMT.

The functional to be extremized is 
\be
S^{\mu}_{,\mu}=\int\;Dp\;\left\{\frac{1}{\tau}I_{col}\left[Z\right]\ln\left[1+Z\right] +\zeta_{\mu\nu}p^{\mu}p^{\nu}f_0Z\right\} -\zeta_{\mu\nu}\Pi^{\mu\nu}
\te
where $\zeta_{\mu\nu}$ are Lagrange multipliers; they must be traceless and transverse. We get the equation

\be
\left\{\ln\left[1+Z\right]+\frac{Z }{1+Z}\right\}=2X+\delta\beta_{\rho}p^{\rho}\mathrm{sign}\left(p^0\right)
\te
where
\be
X=\frac{-\tau}2\frac{\zeta_{\mu\nu} p^{\mu}p^{\nu}}{\left| -u_{\rho}p^{\rho}\right|}
\te
and
\be
\delta\beta_{\rho}=K_{\rho\sigma}\int\;Dp'\;p'^{\sigma}\left(-u_{\mu}p'^{\mu}\right)  f_0\left( p'\right) \ln\left[1+Z\left( p'\right)\right] 
\te
$\delta\beta_{\rho}=0$ to first order.

The point is that this equation has bounded solutions for any value of $X$.  We may also regard it as a means to obtain a formal series solution for $Z$ in powers of $\tau$.
If we keep only the first order term, $\delta\beta_{\rho}=0$ and we get $Z=X$ as in  the Grad ansatz.
We obtain an equation for $\zeta_{\mu\nu}$ by matching moments of the Boltzmann equation \cite{CalPR12,Cal13}.

In this way the EPVM is instrumental in finding a closure valid to all orders in the relaxation time, without  enlarging the number of degrees of freedom beyond the addition of the nonequilibrium tensor $\zeta_{\mu\nu}$, and including the Grad ansatz as its linearized approximation.


\section{Non abelian hydrodynamics}

In last Section we have gone through the derivation of hydrodynamics for a viscous conformal neutral fluid. We now want to generalize this derivation to include color degrees of freedom \cite{CalHu08,manuelrev}.

We are interested in obtaining an effective theory for the dynamics of a system of colored particles interacting with nonabelian classical gauge fields. In this work we will deal with scalar particles coming in three colors, which in our simple model would represent massless and spinless quarks. 
We shall therefore consider a classical Yang-Mills field coupled to conformal scalar matter in the fundamental representation of $SU\left( 3\right) $. 

In what follows, we will use $\mu,\nu,\ldots $ to denote world indices and $a,b,\ldots$ to denote internal (color) indices. We shall denote with $N=3$ the dimension of the fundamental representation, and use $n$ to indicate a generic dimension ($n=3$ or $n=8$ for the fundamental or adjoint representations, respectively). 


The generators $\mathbf{T}_a$ are traceless hermitian $n\times n$ matrices with commutation relations

\be
\left[\mathbf{T}_a,\mathbf{T}_b\right] =iC^c_{ab}\mathbf{T}_c
\label{0}
\te
and trace

\be 
\mathrm{tr}\mathbf{T}_a\mathbf{T}_b=\frac12\delta_{ab}
\label{01}
\te
The Yang-Mills field is $\mathbf{A}_{\mu}=A^a_{\mu}\mathbf{T}_a$. The field tensor 

\be
\mathbf{F}_{\mu\nu}=\partial_{\mu}\mathbf{A}_{\nu}-\partial_{\nu}\mathbf{A}_{\mu}-ig\left[\mathbf{A}_{\mu},\mathbf{A}_{\nu}\right]
\label{1}
\te
belongs to the adjoint representation of the gauge group. 


The equations of motion for the Yang-Mills field are 

\be
\mathbf{D}_{\mu}\mathbf{F}^{\mu\nu}=-\mathbf{P}\left[\mathbf{J}_{\nu}\right]
\label{3}
\te
where the covariant derivative is

\be
\mathbf{D}_{\mu}\mathbf{X}=\partial_{\mu}\mathbf{X}-ig\left[\mathbf{A}_{\mu},\mathbf{X}\right]
\label{4}
\te
$\mathbf{P}$ is a projection operator

\be
\mathbf{P}\left[\mathbf{X}\right]=2\sum_a\mathbf{T}_a\mathrm{tr}\mathbf{T}_a\mathbf{X}=\mathbf{X}-\frac1n\mathrm{tr}\mathbf{X}
\label{4.01}
\te
Eq. (\ref{3}) implies the Bianchi identity

\be
\mathbf{P}\left[\mathbf{D}_{\mu}\mathbf{J}^{\mu}\right]=0
\label{5}
\te
We also have the energy-momentum tensor

\be
T^{\mu\nu}_{YM}=\mathrm{tr}\mathbf{T}^{\mu\nu}_{YM}
\label{6}
\te
where

\be
\mathbf{T}^{\mu\nu}_{YM}=\mathbf{F}^{\mu}_{\lambda}\mathbf{F}^{\nu\lambda}-\frac14g^{\mu\nu}\mathbf{F}^{\lambda\rho}\mathbf{F}_{\lambda\rho}
\label{7}
\te
is traceless in world indices. Using the identity $\mathbf{D}_{\left(\mu\right.}\mathbf{F}_{\left.\nu\lambda\right)}=0$ (where brackets mean symmetrization) we get

\be
\mathrm{tr}\left\{\mathbf{D}_{\mu}\mathbf{T}^{\mu\nu}_{YM}+\mathbf{J}_{\lambda}\mathbf{F}^{\nu\lambda}\right\}=0
\label{8}
\te


The matter fluid is described by the matrix current $\mathbf{J}^{\mu}$ obeying the Bianchi identity (\ref{5}) and by its energy momentum tensor obeying 

\be 
T^{\mu\nu}_{\nu}=-T^{\mu\nu}_{YM;\nu}=
\mathrm{tr}\mathbf{J}_{\lambda}\mathbf{F}^{\mu\lambda}
\label{ecT}
\te 

\subsection{Kinetic theory}

\label{kin}
The kinetic equation which governs the evolution of the one-particle distribution matrix $\mathbf{f}$ reads 

\be
p^{\mu}\left[\mathbf{D}_{\mu}\mathbf{f}-\frac g2\left(\mathbf{F}_{\mu\nu}\frac{\partial\mathbf{f}}{\partial p_{\nu}}+\frac{\partial\mathbf{f}}{\partial p_{\nu}}\mathbf{F}_{\mu\nu}\right)\right]=\mathrm{sign}\left(p^0\right)\mathbf{I}_{col}
\label{9}
\te
where 

\be 
\mathbf{D}_\mu \mathbf{f} = \partial_\mu \mathbf{f} - ig [\mathbf{A}_\mu,\mathbf{f}]
\te
with $\mathbf{A}_\mu$ expressed in the fundamental representation. $\mathbf{f}\left(X,p\right)$  is an $N\times N$ matrix ($N=3$ for quarks) and obeys $\mathbf{f}^{\dagger}\left(X,p\right)=\mathbf{f}\left(X,p\right)$. 


The nonabelian current reads

\be
\mathbf{J}_{\lambda}=g\int\:Dp\;p_{\lambda}P\left[ \mathbf{f}\right] 
\label{10}
\te
where $Dp=d^4p\delta\left(p^2\right)/\left(2\pi\right)^3$.

The matter stress-energy tensor $T^{\mu\nu}_{m}$

\be
T^{\mu\nu}=\mathrm{tr}\mathbf{T}^{\mu\nu}
\label{11}
\te

\be
\mathbf{T}^{\mu\nu}=\int\:Dp\;p^{\mu}p^{\nu}\mathbf{f}
\label{12}
\te
Eqs. (\ref{5}) and (\ref{8}) are identically satisfied provided

\be
\int\:Dp\:\mathrm{sign}\left(p^0\right)\mathrm{tr}\left(\mathbf{T}_a\mathbf{I}_{col}\right)=
\mathrm{tr}\int\:Dp\:\mathrm{sign}\left(p^0\right)p^{\mu}\mathbf{I}_{col}=0
\label{13}
\te
The entropy current is 

\be
S^{\mu}=-\int\:Dp\;p^{\mu}\mathrm{sign}\left(p^0\right)\mathrm{tr}\left\{\mathbf{f}\left( \ln\mathbf{f}-1\right) \right\}
\label{14}
\te
The entropy production 

\be
S^{\mu}_{,\mu}=-\int\:Dp\;\mathrm{tr}\left\{\hat{\phi}\left[ \mathbf{I}_{col}+\mathrm{sign}\left(p^0\right)
\frac g2p^{\mu}\left(\mathbf{F}_{\mu\nu}\frac{\partial\mathbf{f}}{\partial p_{\nu}}+\frac{\partial\mathbf{f}}{\partial p_{\nu}}\mathbf{F}_{\mu\nu}\right)\right]\right\rbrace 
\te 
where
\be
\hat{\phi}=\ln\mathbf{f}
\te
\subsection{Entropy production variational method}
We now consider the derivation of the nonabelian hydrodynamics from the kinetic theory, by applying the EPVM approach. Our first step is to identify the $\beta_{\mu}$ four-vector from the given $T^{\mu\nu}$ and use it to construct an equilibrium (colorless) solution

\be
f_0=\exp\left\{\mathrm{sign}\left(p^0\right)\beta_{\mu}p^{\mu}\right\}
\te
We now parameterize 

\be 
\hat{\phi}=\mathrm{sign}\left(p^0\right)\beta_{\mu}p^{\mu}+\mathbf{Z}
\te 
or else, working to second order in the nonequilibrium correction,

\be 
\mathbf{f}=f_0\left[ 1+\mathbf{Z}+\frac12\mathbf{Z}^2+\ldots\right] 
\te 
The matrix $\mathbf{Z}$ may be written as

\be 
\mathbf{Z}=Z+Z^a\mathbf{T}_a
\te

The form of the entropy production suggests enforcing the second Law exactly by assuming as a collision term

\be
I_{col}=\frac{-1}{\tau}F\left(\beta_{\mu}p^{\mu}\right)f_0\left\{\mathbf{Z} -A^{\lambda}I_{\lambda}-B\mathbf{T}_aI^a\right\}
\te
where

\bea
I_{\lambda}&=&\int\;Dp\;f_0\;p_{\lambda}\mathrm{sign}\left(p^0\right)FZ\nn
I^a&=&\int\;Dp\;f_0\;\mathrm{sign}\left(p^0\right)FZ^a\nn
B&=&\frac1{D_0}\mathrm{sign}\left(p^0\right) \nn
A^{\lambda}&=&\frac1{D_2}\mathrm{sign}\left(p^0\right)\left( u^{\lambda}u^{\sigma}+\Delta^{\lambda\sigma}\right)p_{\sigma}
\tea 
where

\be 
D_k=\int\;Dp\;f_0F\mid-u_{\lambda}p^{\lambda}\mid ^k
\te 
These choices make the collision integral a symmetric operator.
The entropy production to third order in $\mathbf{Z}$. 
\be
S^{\mu}_{,\mu}=-\int\:Dp\;\mathrm{tr}\left\{\mathbf{Z} \mathbf{I}_{col}\right\}
\te
The functional to be minimized is then

\be
\int\:Dp\;\mathrm{tr}\left\{\frac12\mathbf{Z} \mathbf{I}_{col}+\left[g\hat{\zeta}_{\lambda}p^{\lambda}+\zeta_{\lambda\rho}p^{\lambda}p^{\rho}\right]f_0\left[\mathbf{Z}+\frac12\mathbf{Z}^2\right]\right\}
\te
where $\mathrm{tr}\hat{\zeta}_{\lambda}=u^{\lambda}u^{\rho}\zeta_{\lambda\rho}=0$. 


We expand $\mathbf{Z}=\mathbf{Z}^{\left(1\right)}+\mathbf{Z}^{\left(2\right)}$ and similarly the Lagrange multipliers $\hat{\zeta}_{\lambda}$ and $\zeta_{\lambda\rho}$. Then (we write $\hat{\zeta}_{\lambda}^{\left(1\right)}=\hat{\zeta}_{\lambda}$ for short, also for $\zeta_{\lambda\rho}$)

\bea
-\mathbf{I}_{col}\left[\mathbf{Z}^{\left(1\right)}\right]&=&
f_0\left[g\hat{\zeta}_{\lambda}p^{\lambda}+\zeta_{\lambda\rho}p^{\lambda}p^{\rho}\right]\nn
-\mathbf{I}_{col}\left[\mathbf{Z}^{\left(2\right)}\right]&=&f_0\left[\frac g2\left(\hat{\zeta}_{\lambda}\mathbf{Z}^{\left(1\right)}+\mathbf{Z}^{\left(1\right)}\hat{\zeta}_{\lambda}\right)p^{\lambda}\right.\nn
&+&\left.\zeta_{\lambda\rho}p^{\lambda}p^{\rho}\mathbf{Z}^{\left(1\right)}+g\hat{\zeta}_{\lambda}^{\left(2\right)}p^{\lambda}+\zeta_{\lambda\rho}^{\left(2\right)}p^{\lambda}p^{\rho}\right]
\tea
These equations place restrictions on the Lagrange multipliers. To first order $\hat{\zeta}_{\lambda}$ must be transverse, and  $\zeta_{\lambda\rho}$ must be transverse and traceless. 

The solution is
\be
\mathbf{Z}^{\left(1\right)}=\frac{\tau}{F}\left[g\hat{\zeta}_{\lambda}p^{\lambda}+\zeta_{\lambda\rho}p^{\lambda}p^{\rho}\right]
+\mathbf{Z}^{\left(1\right)}_{hom}
\te
$\mathbf{Z}^{\left(1\right)}_{hom}$ is a homogeneous solution. Leaving out a temperature shift, it is

\be 
\mathbf{Z}_{hom} =\frac{D_0}{gA_1}B\mathbf{q}
\te 
for some arbitrary momentum independent  matrix $\mathbf{q}$. 

The first order nonequilibrium current reads

\bea
\mathbf{J}_{\mu}&=&g\int\:Dp\;p_{\mu}f_0\left\lbrace \frac{\tau}{F}g\hat{\zeta}_{\lambda}p^{\lambda}
+\mathrm{sign}\left(p^0\right) \frac{\mathbf{q}}{gd_1}\right\rbrace \nn
&=&\mathbf{q}u^{\mu}+\frac{g^2\tau}{3}F_2\hat{\zeta}_{\mu}
\tea
where

\be 
F_k=\int\;Dp\;\frac{f_0}F\mid-u_{\lambda}p^{\lambda}\mid ^k
\te 
The VEMT reads

\be 
{\tau}^{\mu\nu}=\frac{2\tau F_4}{15}\zeta^{\mu\nu}
\te
To obtain the equations of motion, we substitute our ansatz for the distribution function in the kinetic equation. Keeping only first order terms we get

\be
p^{\mu}\mathbf{D}_{\mu}{f_0}\left[ 1+\mathbf{Z}^{\left(1\right)}\right]  
-g\frac{\partial}{\partial p_{\nu}}p^{\mu}{f_0}\mathbf{F}_{\mu\nu}=
-\mathrm{sign}\left(p^0\right)f_0\left[g\hat{\zeta}_{\lambda}p^{\lambda}
+\zeta_{\lambda\rho}p^{\lambda}p^{\rho}\right]
\te
Separating out the colorless parts we get

\be
p^{\mu}\mathbf{\partial}_{\mu}\left\lbrace {f_0}\left[ 1+\frac{\tau}{F}\zeta_{\lambda\rho}p^{\lambda}p^{\rho}\right]\right\rbrace =
-\mathrm{sign}\left(p^0\right)f_0\zeta_{\lambda\rho}p^{\lambda}p^{\rho}
\te
This is the usual equation from colorless Grad. 

The color part of the kinetic equation yields

\be
p^{\mu}\mathbf{D}_{\mu}{f_0}\left[ \frac{\tau}{F}g\hat{\zeta}_{\lambda}p^{\lambda}
+\frac{\mathrm{sign}\left(p^0\right)}{gd_1}\mathbf{q}\right]  
-g\frac{\partial}{\partial p_{\nu}}p^{\mu}{f_0}\mathbf{F}_{\mu\nu}=
-\mathrm{sign}\left(p^0\right)f_0g\hat{\zeta}_{\lambda}p^{\lambda}
\te
From the second moments of this equation we get, in the rest frame

\be
\hat{\zeta}_{i}=\frac{2A_2}{A_3}\mathbf{E}_{i}-\frac{1}{g^2A_1}\mathbf{D}_{i}\mathbf{q}-\frac{\tau F_4}{A_3}\mathbf{D}_{0}\hat{\zeta}_{i}
\te
Which is a Maxwell - Cattaneo type equation.

\section{Conclusions}
To summarize, we have demonstrated the construction of hydrodynamics for a viscous fluid carrying color charges. This theory has, besides the inverse temperature four vector which characterizes the color neutral equilibrium states, new degrees of freedom, namely the color fugacities $\mathbf{q}$ and the Lagrange multipliers 
$\hat{\zeta}_{i}$ and $\zeta_{\lambda\rho}$. The dynamics of the former is determined by charge conservation, while the latter obey Maxwell - Cattaneo type equations. In the limit where the relaxation time $\tau\to 0$, $\hat{\zeta}_{i}$ reduces to the sum of two terms, one proportional to the chromoelectric field (Ohm's law) and the other to the fugacity gradient (Fick's law), while the VEMT reduces to the usual shear viscosity term.

The two basic problems where such a formalism may become useful are the study of chromomagnetic instabilities \cite{inst}  and of energy deposition by a fast particle \cite{jets}. The former problem is relevant to the study of the early times of the collision, as those instabilities may act to speed up isotropization of the plasma, while the latter is relevant to the analysis of the so-called jet quenching. Work is underway in both directions, and we look forward to report on it at future meetings.

\section*{Acknowledgment}
This work has been developed in collaboration with Jer\'onimo Peralta Ramos. It is supported in part by Universidad de Buenos Aires, CONICET and ANPCYT (Argentina)

\end{document}